\documentclass[10pt, onecolumn,floatfix]{revtex4}
\usepackage{layout}
\usepackage{latexsym}
\usepackage{amsmath,amsthm,amsfonts,amssymb,amscd}
\usepackage{graphicx}
\usepackage{textcomp}

\begin{document}

\title{High resolution experimental parameter space of a chaotic circuit}
\author{Francisco F. G. de Sousa$^{1,2}$, Rero M. Rubinger$^1$, Jos\'e C. Sartorelli$^3$, Holokx A. Albuquerque$^4$, and Murilo S. Baptista$^5$}

\affiliation{$^1$Instituto de F\'{i}sica e Qu\'{i}mica, Universidade Federal de Itajub\'a, Itajub\'a, MG, Brazil, $^2$Instituto Federal do Sul de Minas, Inconfidentes, MG, Brazil, $^3$Instituto de F\'{i}sica, Universidade de S\~ao Paulo, S\~ao Paulo, SP, Brazil, $^4$Departamento de F\'{i}sica, Universidade do Estado de Santa Catarina, Joinville, SC, Brazil, $^5$Institute of Complex Systems and Mathematical Biology, University of Aberdeen, SUPA, Aberdeen, UK.}

\begin{abstract}
This work shows how to create  reliable, autonomously, and reproducible high resolution parameter spaces of nonlinear systems. To test our scientific and technical approach we show that complex features observed in the numerically obtained parameter space of the Chua's circuit  can be reproduced experimentally, such as the period-adding bifurcation route describing periodic regions whose size decrease exponentially with their period. Consequently, indicating that periodic behaviour with higher period is unlikely to be observed. The high-resolution span of parameters was possible by the use of a newly designed potentiometer, which can have its value autonomously altered. To have such resistances we developed in series arrays of resistors short-circuited by relays as discrete potentiometers with 1024 steps, and resolutions of 0.100 $\Omega $  for $r_{L}$ in series with the inductor,
and 0.200 $\Omega $ for R connecting the two capacitors. 
\end{abstract}

\maketitle

\section{Introduction}
      So far, only a few chaotic circuits had their parameter spaces experimentally obtained, a space that contains information about the Lyapunov exponents and the periodicity of the observed time series for a range of parameters \cite{Maran, Viana, Viana2, Sack, Stoop, Stoop2}. The relevance of studying parameter spaces of nonlinear systems is that it allows us to understand how periodic behaviour, chaos and bifurcations come about in a nonlinear system. In fact, parameters leading to the different behaviours are strongly correlated. Chaotic and periodic regions appear side by side in all scales in universal shapes and forms. Not long ago, Gallas \cite{Gallas} numerically observed periodic structures embedded in parameter chaotic regions, in the parameter space of the H\'enon map. He coined a name for them: shrimps; given the remarkable similarity between parameter regions leading to periodic behaviour, which appear all along special directions, and how shrimps in barbecue stick are so well organised. For some classes of nonlinear systems such as the one studied here the periodic structures appear aligned along spiral curves describing parameters for saddle-node bifurcations and super-stable behaviour, and that cross transversally parameter curves containing homoclinic bifurcations \cite{rene, Avila, Vitolo, barrio, Cabeza, Albu}.
      
			Experimentally, the difficulty in obtaining parameter spaces resides in a reliable method to vary precisely a parameter, usually a resistance, in a controlled, autonomously, and reproducible fashion. Parameters are not constant and suffer time varying alterations. This factor becomes even more severe when the experiment is done over long time spans. Finally, the numerical resolution of the parameters and their nominal values cannot be achieved or reproduced experimentally. Even very simple nonlinear electronic systems cannot have their behaviours reproduced numerically; the main reason is that the electronic components have non-ideal characteristic curves. In particular, and with regard to this work, an idealised piece-wise linear component behaves non-linearly in an experiment. These factors are behind our motivation to propose an approach to obtain experimental parameter spaces that are not only reliable, autonomous, and reproducible, but that can also reproduce nominal parameter values considered in numerical experiments. High resolution parameter space allows one to reproduce experimentally the self-similar topological character observed in numerically obtained parameter spaces. 
			
			In order to tackle these experimental issues, previous works have proposed different strategies to construct experimental parameter spaces. Maranh\~ao et. al. considered a manual or step motor control of precision multi-turn potentiometer \cite{Maran}. Stoop et. al. \cite{Stoop,Stoop2} used a proto-board to do the experiment and varied by hand 490 x 162 values of a negative resistor and an inductor to produce parameter spaces. In Ref. \cite{Viana}, it was used a Keithley power source controlled by LabView®as one parameter and a precision potentiometer manually controlled as the second parameter. The use of a digital potentiometer would provide a reliable, autonomously, and reproducible way to obtain parameter spaces. However, the available digital potentiometers on the market are limited to resistances above 1 k$\Omega$  and usually not more than 256 steps are possible. In addition, their control is not easy to carry out with typical data acquisition systems or LabView®. 
			
			The technological novelty in this work is to present the design of a digital potentiometer with precisely calibrated small resistance steps (as low as 0.100 $\Omega$) that allows 1024 steps (or even more) to change the resistance. With these potentiometers and a set of resistors, switches and relays, we were able to autonomously obtain a high resolution parameter space of the Chua's circuit  (with resolutions of 400 x 562 and 1023 x 126 points, varying one resistor with step sizes of 0.2$\Omega$ and another with 0.1$\Omega$), which could remarkably reproduce the numerically obtained parameter spaces. 
			
    This proposed electronic component allowed us to carry out a detailed experimental investigation of the parameter space of a modified Chua's circuit, namely we have characterised this circuit by varying the resistance (R) and the inductor resistance ($r_{L}$) \cite{Torres,Albu2}. Among other accomplishments, we have demonstrated that even higher period periodic windows and the complex topological structure of the scenario for the appearance of periodic behaviours can be observed experimentally. Simulations, considering a normalized equation set that models the Chua's circuit, and also the normalised version of the experimental i(V) curve, were carried out in order to demonstrate that the occurrence of periodic structures observed in the high-resolution experimental parameter space could also be numerically observed. In particular, we showed numerically and experimentally that this parameter space presents self-similar periodic structures, the shrimps, embedded in a domain of chaos \cite{Maran,Viana,Stoop,Viana2,Stoop2,Sack,Gallas,Avila,Vitolo,Cabeza,Albu}. We also show experimentally that those self-similar periodic regions organize themselves in period-adding bifurcation cascades, and whose sizes decrease exponentially as their period grows \cite{Maran}. We also report on malformed shrimps on the experimental parameter space, result of tiny nonlinear deviations from a symmetric piecewise linear i(V) curve. 
    
We have considered the Chua's circuit \cite{Chua} to perform our study experimentally and numerically because this circuit has been studied in many applications such as in chaos control \cite{Wu,Campos}, synchronisation \cite{Zhang,Zhang2} and others, but the use of a higher precision potentiometers here proposed can be used to characterise, study, and precisely control the behaviour of any electronic equipment.  Detailed information about the design of this new electronic component is being provided in this manuscript.  
    
\section{Experimental and Numerical Aspects}
\begin{figure}[h]
\includegraphics[width=3.4in]{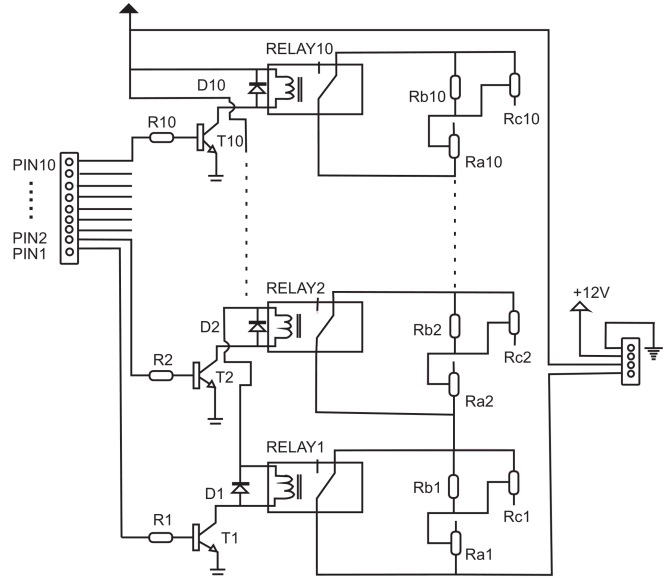}
\caption{Schematics of the designed adjustable digital potentiometer. Only three of ten resistor circuits of the in series association are shown in order to clearly present their components. See text for component and respective function description.}
\label{Fig1}
\end{figure}

    Prior to the development of the chaotic circuit, we designed the digital potentiometer with the aim of obtaining high-resolution parameter spaces considering precise and tiny variations in the resistances, which allows reproducible experiments happening for the same parameters. The digital potentiometer is presented in Fig. \ref{Fig1}. Only three resistor circuits of the in series association were presented in the figure in order to better illustrate its structure. The 10 pin left connector stands for the digital data coming from input/output (I/O) digital ports of the DAQ board. The pins and all other components have their index identified with a number X in the end (e.g. PinX). PinX is associated to the bit number numerated from the least significant digit. Each pin is connected to a transistor TX in a switching configuration with RX as a resistor to limit base current, the collector connected to the anode of a diode DX and one pole of the coil of the relay. The cathode of DX is connected to the +12V battery pin and the other pole of the coil. All the TX emitters are connected to the ground pin of the battery. All relays are connected in parallel with a 3 resistor circuit. In the non-switched position, relays short-circuit the resistor circuit X, leading to a negligible in series resistance for the branch X of the circuit. When switched, they open the short circuit and RaX, RbX and RcX equivalent resistance add to the in series resistor network. RaX and RcX are calibration resistances while RbX is a precision resistor. If RbX is less than the necessary RaX is a multi-turn trimpot to add a small resistance. If the RbX is more than the desired the parallel resistor RcX is used to lessen the equivalent value and RaX is substituted by a short circuit. 
The digital potentiometer idea was captured from the structure of a switched resistor digital to analog converter which contains a parallel resistor network. As in the case of the circuit, one 60 Ah 12.0 V car battery was used to drive the digital potentiometers. The calibration was done using a Keithley digital multimeter model 2001 in the four wire mode, i.e. in order to subtract leads resistance. The relays of the series association were switched by transistor driver circuits connected to I/O digital ports of the data acquisition board used for control and data acquisition. Thus it was possible to write a ten digit binary number in order to select one of the 1024 possible combinations. Then the potentiometer resistances R and $r_{L}$ follows the equation:
\begin{equation}
\centering {= step*(bit0*2^0+bit1*2^1+...+bit9*2^9).}
\label{eq1}
\end{equation}

\begin{equation}
i_{d}(x)=\left\{ 
\begin{array}{cc}
 -32.51240-4.766x & x < -5.430, \\ 
x-0.82999 & -5.430\leq x < -1.000, \\ 
1.84957x & \left\vert x\right\vert \leq 1.000, \\ 
x+0.86378 & 1.000 < x \leq 5.929, \\ 
37.35590-5.152x & x > 5.929,
\end{array}
\right.   \label{eq2}
\end{equation}

\begin{figure}[h]
\includegraphics[width=3.4in]{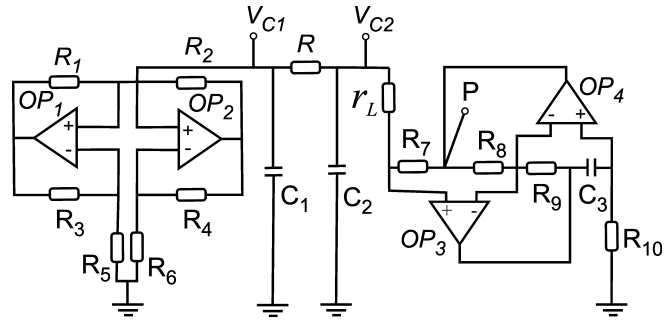}
\caption{Chua's Circuit using the electronic inductor and indicating the measuring points x, y and the P point. The current through the inductor is defined as $I_{L} = (V_{P}-V_{C2})/R_{7}$.}
\label{Fig2}
\end{figure}

\begin{figure}[h]
\includegraphics[width=3.4in]{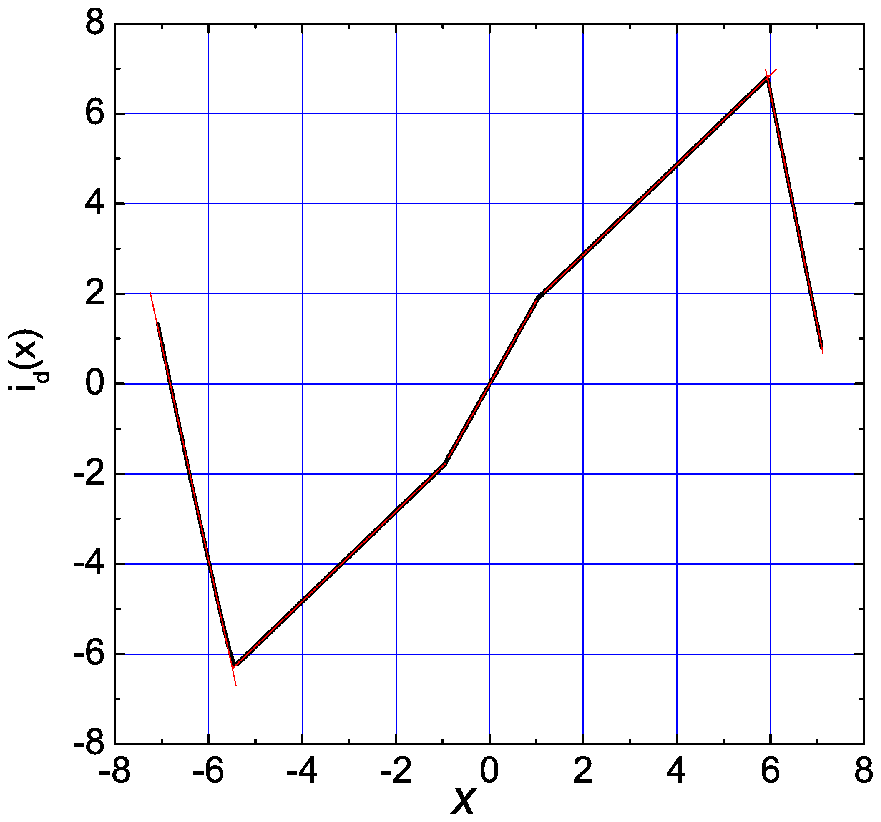}
\caption{ (Color online) $i_{d}(x)$ characteristic of the circuit presented in Fig. \ref{Fig2}. The linear fittings are presented in the figure. The equations corresponding to linear fittings are presented in the experimental section. The normalization considered equations $x = V_{C1}/B_{P}$ and $i_{d}(x)=i(x)/(m_{o}/B_{P})$ and the parameters $m_{o} = -4.156315$ mS and $B_{P}= 1.3850075$ V. Insets shows the two external kink regions in detail.}
\label{Fig3}
\end{figure}

	For R we used two-step values: 0.200 $\Omega$ and 2.000 $\Omega$ and for $r_{L}$ just one value for step, i.e. 0.100 $\Omega$.
    Chua's circuit scheme is presented in Fig. \ref{Fig2}, constructed in a single face circuit board with the same scheme of \cite{Rero} i.e., $C_{1} = 23.50$ nF, $C_{2} = 235.0$ nF, and $L = 42.30$ mH. These values were obtained from the combination of passive commercial available components and measured with a Keithley digital multimeter model 2001 or an impedance analyser for the reactive components. The measurement of components allowed the choice of the closest possible values to components, better than the factory precision. We evaluate the oscillation main frequency as a rough approximation by $1/(2\pi(LC_{2})^{1/2})$ which gives 1596 Hz. Further increase on frequency, i.e. by reducing passive component values, seems to destroy periodic structures that are observed in this circuit. Thus this oscillation frequency was the best choice that allowed us to store large time series for data analysis. Other parameters were experimentally determined. The five-fold piecewise linear element consists of two operational amplifiers (OPAMP) and resistances $R_{1}$to $R_{6}$, the $i(V_{C1})$ characteristics was defined and normalised by the scheme $x = V_{C1}/B_{P}$ and $i_{d}(x)=i(x)/(m_{0}B_{P})$ with $m_{0}=4.156315$ mS and $B_{P} = 1.38501$ V. Here S stands for the inverse resistance unity. This curve is presented in Fig. \ref{Fig3}, with 5-fold linear fittings used for simulations given by:
	

with the significant digits limited by the fitting accuracy. The circuit variables $V_{C1}$, $V_{C2}$ measuring points are indicated by their respective probes. They correspond to the voltages across the capacitors $C_{1}$ and $C_{2}$ and the third variable, the current $I_{L}$, is obtained from the relation $I_{L} = (V_{P}-V_{C2})/R_{7}$ where $V_{P}$ is the voltage indicated by the P probe. The electronic inductor is defined by two OPAMPs connected to the resistors $R_{7}$, $R_{8}$, $R_{9}$, $R_{10}$ and $r_{L}$ together with the capacitor $C_{3}$. This is a gyrator circuit with inductance as $L = (C_{3}R_{7}R_{9}R_{10})/R_{8}$. The resistor R, used as one control parameter, was composed by a 1193.445 $\Omega$ resistor in series with the digital potentiometer that varied in the range of 1193.445 $\Omega$ to 1993.445 $\Omega$ with a 0.200 $\Omega$ and, in a second experiment, the potentiometer was varied in the range of 1414.642 $\Omega$ to 1619.242   $\Omega$ a 0.200 $\Omega$. The other control parameter, the resistor $r_{L}$ was a digital potentiometer whose range varied within the interval [8.077 64.277] $\Omega$ in the first experiment and [28.077 40.677] $\Omega$ in the second experiment, both cases with a 0.100 $\Omega$ stepsize. The Chua's circuit, built with TL084 Operational Amplifiers (OPAMP) was fed by two 12.0 V, 7 Ah no-break batteries and the voltage across $C_{1}$, passed by a simple OPAMP buffer, was measured by a national instruments data acquisition (DAQ) interface, model PCI-6259 with 16 bit resolution, maximum sampling rate of 1.25 Msamples/s, for data storage. Also, LabView® was used to data acquisition and analysis \cite{Viana, Viana2}. A Keithley 2400 voltage/current source in series with the Chua's diode was applied to obtain the i(V) data. For each time series, the potentiometers R and $r_{L}$ were switched by a LabView® routine with values previously determined and calibrated to give precise equivalent steps. After calibration, the 1024 values of each potentiometer were tested with a linear fitting, giving slopes equal to steps up to four significant digits. 
The 5-fold piecewise linear $i_{d}(x)$ characteristics were presented in Fig.  \ref{Fig3} and Eq. (2). We have carried out experiments with the Chua's circuit recollecting time series for the calculation of the Lyapunov exponents. Time series were generated with a 50 Ksamples/s and a 7 s length. As transient, after setting the pair of parameters R and $r_{L}$, a 50 s waiting time was considered. 

It is worth commenting that data acquisition to obtain 562 x 400 meshes of time series lasted 3 weeks. The fact that the experimental circuit could reproduce many relevant structures obtained numerically implies that the general bifurcation scenario of these periodic windows of the parameter space are robust to external perturbations and should be expected to be observed in nature. During this time the batteries were recharged many times, and the room temperature has varied during a 24h period and during these 3 weeks. The circuit is not switched off between subsequent measurements only at the highest R when recharge was necessary. Each restart of the system has changed the attractor initial conditions. However, because the parameters were varied from high R to lower R values and from low $r_{L}$ to higher $r_{L}$ values, the trajectory at restart was always going towards the same fixed points or simple periodic orbits. By starting at parameters leading to such attractors provide results as if the system was never switched off. Thus, this form of covering the parameter space allows best reproducibility since the next attractor is equivalent of numerical continuation, i.e. the last state of the system at a pair of control parameters is the beginning of the next one.

\begin{figure}[h]
\includegraphics[width=3.4in]{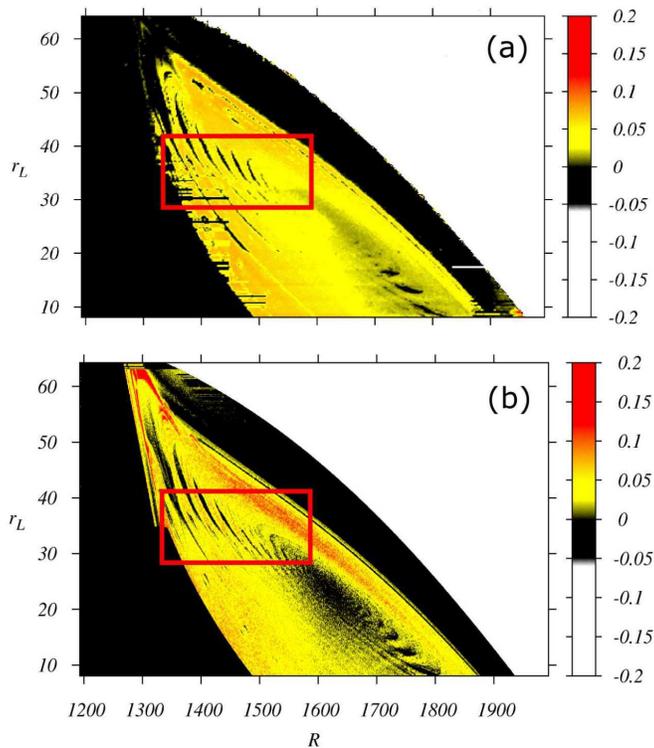}
\caption{(Color online) Parameter spaces of the Chua's circuit. White to black stands for periodic orbits and for fixed points, yellow to red color for chaotic orbits. (a) Experimental parameter space diagram associating color scale to $\lambda$. Resolution of parameters R and $r_L$ is 2$\Omega$, and we have considered a mesh with 400 values for R and 562 values for $r_L$.   (b) Corresponding simulated parameter space obtained from using the model of Ref. \cite{Albu2} but with a 5-fold piece-wise $i_{d}(x)$ Eq. (2). Resolution of parameters R and $r_L$ is 0.5$\Omega$ and 0.1$\Omega$, respectively, and we have considered a mesh with 1,600 values for R and 700 values for $r_L$. }
\label{Fig4}
\end{figure}

Experimental characteristic curves of the Chua's circuit are often asymmetric. However, some authors have done simulations considering a symmetric piecewise-linear function. In this work, successful reproduction of the experimental parameter spaces is also a consequence of the fact that our correspondent  simulations were performed considering the non-symmetric $i_{d}(x)$ [in Eq. (2)] to integrate Chua's differential equations. In this work, we have considered the same set of differential equations and normalized parameters presented in full detail in Ref. \cite{Albu2}. 

\section{Results and discussion}

The parameter space in Fig. \ref{Fig4}(a) shows by colours the values of the largest Lyapunov exponent $\lambda$, calculated by the method of Sano and Sawada \cite{Sano} from the $400 \times 562$ experimental time series with R and $r_{L}$ as the control parameters.  The simulated parameter space in Fig. \ref{Fig4}(b)  considered also the values of $\lambda$ obtained from time series generated by 1,600  values of R and 700 values of $r_{L}$ in the same range of the experimental data. In the case of simulations, $\lambda$
was obtained from the tangent space method, and the values are in units of integration step. This form of calculating $\lambda $
allows faster simulations but produce absolute values of the exponents distinct from the corresponding experimental values. The time units of the experimental Lyapunov exponent was rescaled to match the exponents from numerical data. 
    
A larger range parameter space, i.e. with R from 1193.445 $\Omega$ to 1993.445 $\Omega$ with 2.000 $\Omega$ step variation and $r_{L}$ from 8.077 $\Omega$ to 64.277 $\Omega$ is presented in Fig. \ref{Fig4}(a). In Fig. \ref{Fig4}(b), we present the correspondent simulations, generated with a 0.100 $\Omega$ step, as described in section 2.	The color scales were defined for the experimental data with a smooth color variation from white to black to represent the range of values of $\lambda$, with $\lambda < 0$,  corresponding to fixed point and periodic attractors, and from black to red to represent the range of values $\lambda \in [0,0.2]$. The transition between periodic to chaotic orbits occurring through saddle-node bifurcations or through a period doubling cascade is characterised by the shift of colours between yellow and orange. In both parameter spaces of Fig. \ref{Fig4}, it is possible to identify the occurrence of complex periodic structures embedded in a chaotic domain and organised in a spiral structure \cite{Gallas,Cabeza,Cardoso}. 
    
\begin{figure}[h]
\includegraphics[width=3.4in]{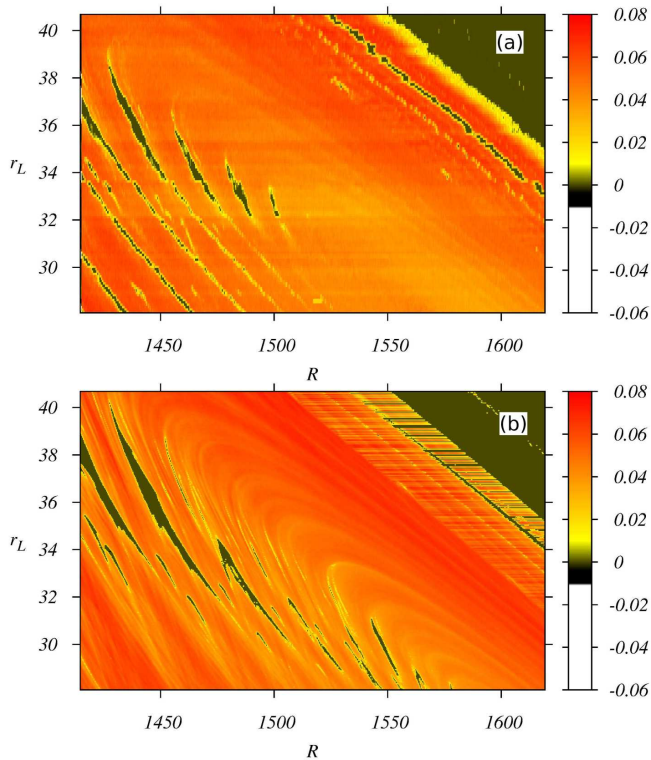}
\caption{(Color online) Magnification of Figs. 4(a) and 4(b). Color scheme is the same used in Fig. \ref{Fig4}. In (a) experimental data with ten times the R resolution compared to that used in Fig. \ref{Fig4}(a). In (b), a 600 $\times$ 600 mesh parameter space with the same high resolution of the parameter R.}
\label{Fig5}
\end{figure}

\begin{figure}[htb]
\includegraphics[width=3.4in]{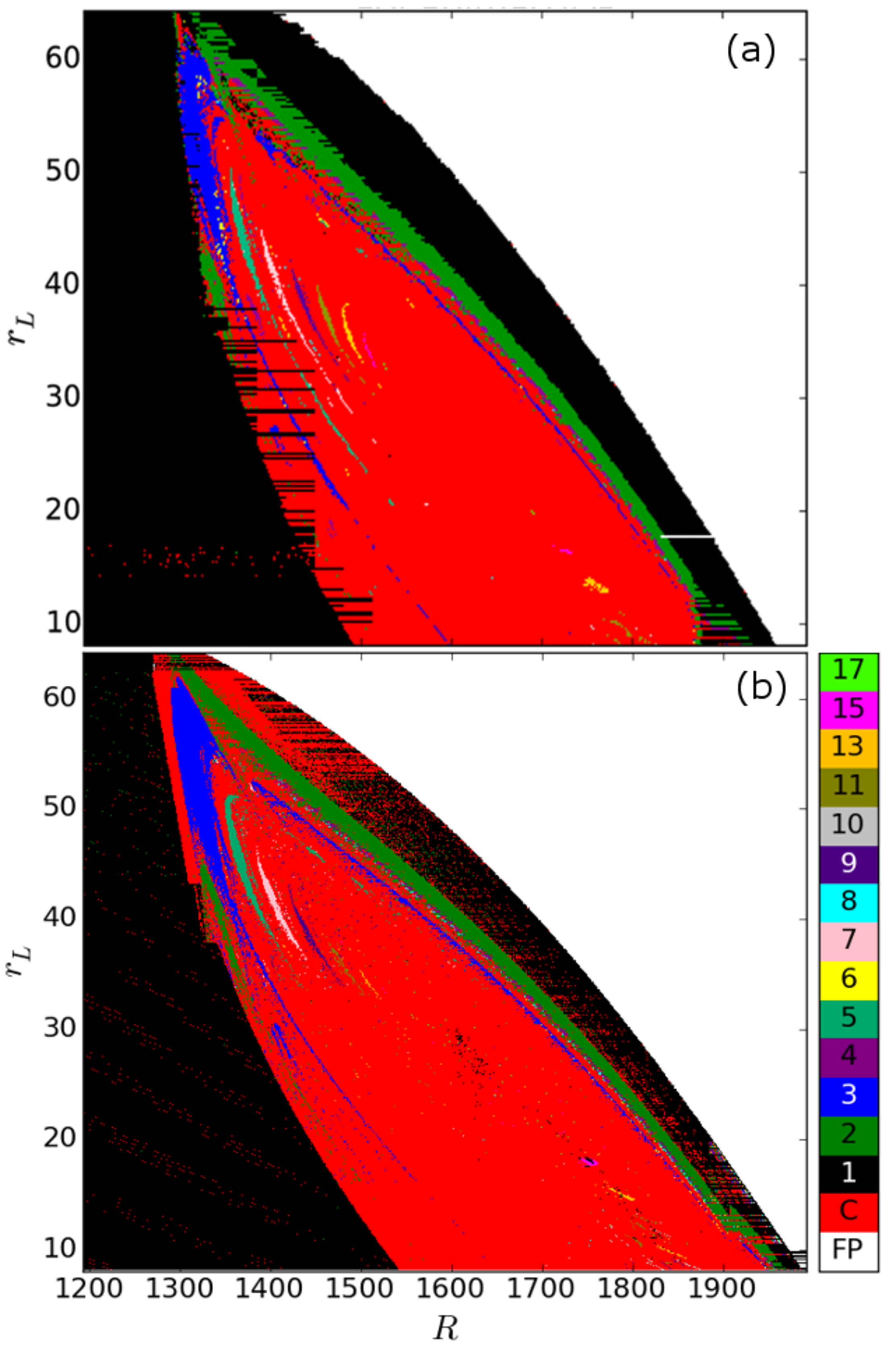}
\caption{(Color online) Periodicity parameter spaces for the parameter spaces of Figs. \ref{Fig4}(a) and \ref{Fig4}(b). Color code for the period of the attractor is presented in the right-hand side band. Notice an odd period-adding bifurcation cascade initiating at the top left corner and heading towards the center of the spiral.}
\label{Fig6}
\end{figure}

\begin{figure}[htb]

\includegraphics[width=3.4in]{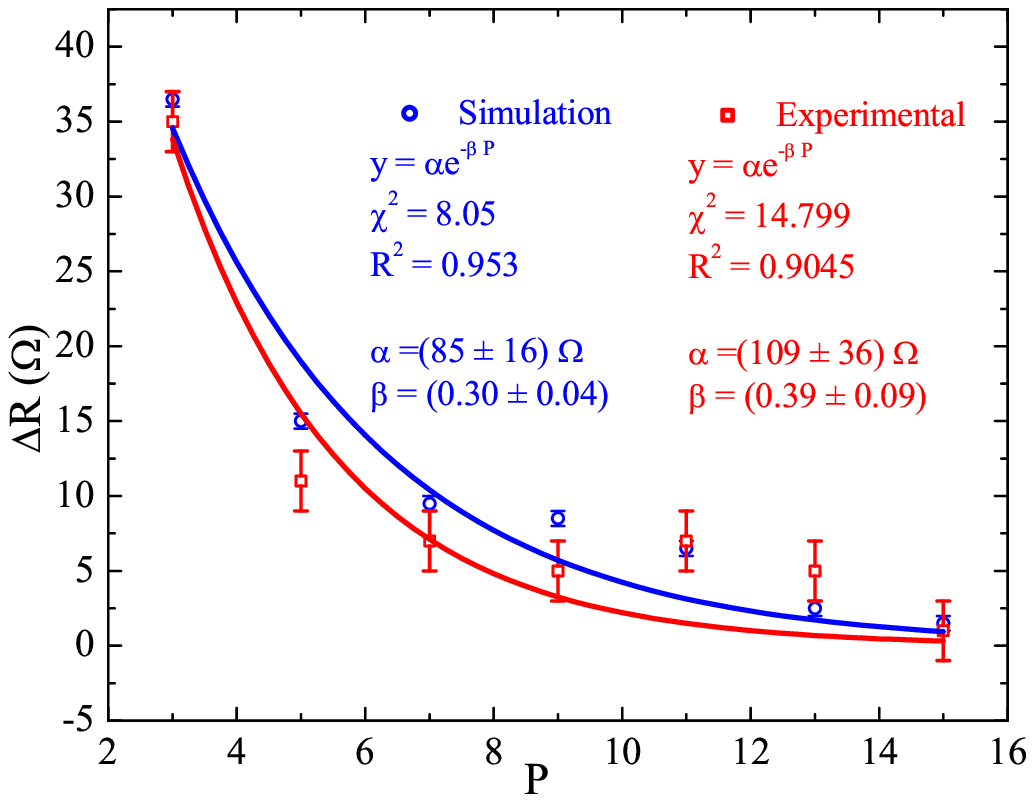} \label{Fig7}

\caption{(Color online) Fitting of the periodic window largest width $\Delta R$ with respect to the attractor period $P$, considering the exponential scaling of Eq. (\ref{eqR}). We considered the error bars of 2 $\Omega$ in the experimental calculations and 0.5$\Omega$ in the simulated ones.}
\end{figure}

Figure \ref{Fig5} shows a high-resolution amplifications of the inner region of the red boxes in Fig. \ref{Fig4} . As in Fig. \ref{Fig4}, Fig. \ref{Fig5}(a) stands for experimental data, obtained with a 0.200 $\Omega$ resolution in R, i.e. ten times more points for this parameter. In Fig. \ref{Fig5}(b), we present a parameter space constructed considering a 600  $\times$  600 mesh of values for R and $r_{L}$ for the corresponding simulations. Notice that both figures are remarkably similar, showing the same complex features. In particular, both figures show details of the complex self-similar organisation of periodic regions.

Despite of all care with the use of batteries instead of switched power sources, the use of a PCB instead of the proto-board, the use of a large metal plate as ground reference to avoid ground-loops, an OPAMP buffer and a 16 bit DAQ board, 
disparities between the experimental and numerical parameter spaces of Figs. \ref{Fig4} and  of Figs. \ref{Fig5} can be noticed. These effects, of experimental nature, are a noise in the color distribution of the parameter space that makes the boundaries to chaos of the complex periodic windows distorted and the presence of horizontal stripes with distinct shades of colours in the chaotic region. The first one is associated with thermal, electrical noise and the unavoidable analog to digital conversion noise. We estimated from the measured time series that this noise is between 1 mV and 2 mV. This represents three to six times the least significant bits of the DAQ board which is close to the minimum noise experimentally realisable. The second effect is associated with temperature and initial conditions. Between consecutive days, the temperature may vary a few $^{o}C$
 and this change the values of resistances and diode voltage of transistors inside the OPAMPs. We have experienced that the changes are minimum in the component values for a $10^{o}C$
 change. We were not able to determine whether the temperature change or the change in initial conditions influence the formation of horizontal stripes in the experimental parameter space. There can exist another source of dissimilarities.  
That the experimental $i_{d}(x)$ curve has a non-linear component at the inflection points such as the local minima within $X \in [-5,-6]$, in contrast to the piecewise linear $i_d(x)$ considered for numerical simulations as in Eq. (\ref{eq2}). 
 
We have also studied the topological properties of the periodic windows appearing in the parameter spaces of Figs. \ref{Fig4}(a)-(b). For that goal, we created the periodicity parameter spaces in Fig. \ref{Fig6}(a), for the experimental time-series, and in Fig. \ref{Fig6}(b), for numerical time-series, indicating by colours the period of the periodic attractors observed in Figs.  \ref{Fig4}(a)-(b). 
In Fig. \ref{Fig7}, we verify the existence of expected scaling laws for the largest width of the periodic windows $\Delta R$ as a function of the period $P$ of the corresponding attractors in the parameter spaces of Fig. \ref{Fig7}(a)-(b), by fitting the following scaling 
\begin{equation}
					\centering {\Delta R=\alpha e^{-\beta P},}
\label{eqR}
\end{equation}
with $\alpha$ and $\beta$ representing fitting parameters. This result led us to conclude that our experimentally and numerically obtained periodic windows have a complex structure as expected in Refs. \cite{Maran,Viana,Lu}. It can be seen from Fig. \ref{Fig6} that periodic windows organize themselves in an odd period adding bifurcation cascade starting from the top left corner of the parameter space towards the spiral center, and whose attractors present period varying from 3 to 17. Window sizes reduce as the period increases and their width $\Delta R$ in $\Omega$ reaches its minimal value of 2 $\Omega$ for periods 15 and 17. 
 Fitting results indicate that the decay exponent is $\beta =0.3 \pm 0.04$ (experimental results) and $\beta=0.39 \pm 0.09$ (numerical results) is in the same order of magnitude of the largest positive Lyapunov exponent of the chaotic attractor in the chaotic regions surrounding the shrimps (about 0.2), as expected \cite{Maran}. 

\section{Conclusions}
    In conclusion, we have successfully built an autonomous, reliable, and reproducible digital potentiometer that allowed precise scientific measurements of physical invariants of a chaotic circuit. 
  As an application of the power of our proposed potentiometer, we reported a high-resolution experimental parameter space of the electronic inductor Chua's circuit that is remarkably similar to the one obtained numerically.  Two of these specially designed digital potentiometers were used to vary the circuit main resistances and the inductor-resistance. 
We showed that, as in the simulations, the experimental parameter space also presents stability islands embedded in a domain of chaos, the shrimps, whose sizes decay exponentially with the period of the attractor.   

\begin{acknowledgments}
The authors thank Prof. Iber\^e Luiz Caldas for the suggestions and encouragement.
The authors FFGS, RMR, JCS and HAA acknowledge the Brazilian agency CNPq and
state agencies FAPEMIG, FAPESP, and FAPESC;  and MSB also acknowledges a EPSRC grant EP/I032606/1.\\
\\
Corresponding author: sartorelli@if.usp.br
\end{acknowledgments}


\begin{thebibliography}{99}
\bibitem{Maran} D. M. Maranh\~ao, M. S. Baptista, J. C. Sartorelli, I. L. Caldas,   Phys. Rev. E \textbf{77}, 037202 (2008).
\bibitem{Viana} E.R. Jr Viana, R.M. Rubinger, H.A. Albuquerque, A.G. de Oliveira, G.M. Ribeiro, Chaos \textbf{20}, 023110 (2010);.
\bibitem{Stoop} R. Stoop, P. Benner, and Y. Uwate, Phys. Rev. Lett. \textbf{105},  074102, (2010).
\bibitem{Viana2} E.R. Jr Viana, R.M. Rubinger, H.A. Albuquerque, F.O. Dias, A.G. de Oliveira, G.M. Ribeiro, Nonlinear Dynamics \textbf{67}, 385-392, 2012.
\bibitem{Stoop2} R. Stoop, S. Martignoli, P. Benner, R. L. Stoop and Y. Uwate, Int. J. Bifurcation Chaos Appl. Sci. Eng.  \textbf{22}, 1230032, 2012.
\bibitem{Sack} A. Sack, J. G. Freire, E. Lindberg, T. Pöschel, and J. A. C. Gallas, Scientific Reports \textbf{3}, 3350, 2013.
\bibitem{Gallas} J. A. C. Gallas, Phys. Rev. Lett. \textbf{70}, 2714-2717 (1993).
\bibitem{Avila} G. M. Ram\'{i}rez-Ávila, J. A. C. Gallas, Phys. Lett. A \textbf{375}, 143-148 (2010). 
\bibitem{Vitolo} R. Vitolo, P. Glendinning, and J. A. C. Gallas, Phys. Rev. E \textbf{84}, 016216 (2011).
\bibitem{rene} R. O. Medrano-T and I. L. Caldas, arXiv:1012.2241 (2010). 
\bibitem{barrio} R. Barrio, F. Blesa, S. Serrano, and A. Shilnikov, Phys. Rev. E {\textbf 84},, 035201 (2011). 
\bibitem{Cabeza} C. Cabeza, C. A. Briozzo, R. Garcia, J. G. Freire, A. C. Marti, J. A. C. Gallas, Chaos, Solitons \& Fractals \textbf{52}, 59-65 (2013).
\bibitem{Albu} H. A. Albuquerque, P. C. Rech, International Journal of Circuit Theory and Applications \textbf{40}, 189-194 (2012).
\bibitem{Chua} L. O. Chua, Journal of Circuits, Systems and Computers \textbf{4}, 117-159 (1994).
\bibitem{Wu} T. Wu, C. Min-Shin, Physica D \textbf{164}, 53-58 (2002).
\bibitem{Campos} C. D. Campos, R. M. Palhares, E. M. A. M. Mendes, L. A. B. Torres and L. A. Mozelli, Int. J. Bifurcation Chaos Appl. Sci. Eng.  \textbf{17}, 3199-3209 (2007).
\bibitem{Zhang} J. Zhang, C. Li, H. Zhang, J. Yu, Chaos Solitons \& Fractals \textbf{21}, 1183-1193 (2004).
\bibitem{Zhang2} Y. Zhang, J. Sun, Phys. Lett. A \textbf{330}, 442-447 (2004).
\bibitem{Torres} L. A. Tôrres, L. A. Aguirre, Electronics Letters \textbf{36}, 1915-1916 (2000).
\bibitem{Albu2} H. A. Albuquerque, R. M. Rubinger, P. C. Rech, Physica D \textbf{233}, 66-72 (2007).
\bibitem{Rero} R. M. Rubinger, A.W. M. Nascimento, L. F. Mello, C. P. L. Rubinger, N. Manzanares Filho, and H. A. Albuquerque, Mathematical Problems in Engineering \textbf{2007} 83893-83909 (2007).
\bibitem{Sano} M. Sano and Y. Sawada, Phys. Rev. Lett. \textbf{55}, 1082-1085 (1985).
\bibitem{Cardoso} J. C. D. Cardoso, H. A. Albuquerque, R. M. Rubinger, Phys. Lett. A \textbf{373}, 2050-2053 (2009).
\bibitem{Lu} J. H. Lu, G. R. Chen, D. Z. Cheng, Int. J. Bifurcation Chaos Appl. Sci. Eng. , \textbf{14}, 1507-1537 (2004).

\end{thebibliography}
\end{document}